\documentclass[12pt,twoside,a4paper]{article}
\usepackage{graphicx}
\usepackage{cite}
\usepackage{url}
\usepackage{color}

\begin{document}
\thispagestyle{empty} 
\title{
\vskip-3cm
{\baselineskip14pt
\centerline{\normalsize DESY 15-178 \hfill ISSN 0418--9833}
\centerline{\normalsize MITP/15--077 \hfill} 
\centerline{\normalsize September 2015 \hfill}} 
\vskip1.5cm
\boldmath
{\bf Inclusive photoproduction of bottom quarks 
for low and medium $p_T$ in the}
\\
{\bf general-mass variable-flavour-number scheme}
\unboldmath
\author{
G.~Kramer$^1$, 
and H.~Spiesberger$^2$
\vspace{2mm} 
\\
\normalsize{
  $^1$ II. Institut f\"ur Theoretische
  Physik, Universit\"at Hamburg,
} 
\\ 
\normalsize{
  Luruper Chaussee 149, D-22761 Hamburg, Germany
} 
\vspace{2mm} 
\\
\normalsize{
  $^3$ PRISMA Cluster of Excellence, Institut f\"ur Physik,
}\\
\normalsize{
  Johannes Gutenberg-Universit\"at, 55099 Mainz, Germany,
}\\ 
\normalsize{
  and Centre for Theoretical and Mathematical Physics and 
  Department of Physics,
}\\
\normalsize{ 
  University of Cape Town, Rondebosch 7700, South Africa
} \vspace{8mm} \\
}
}

\date{}

\maketitle

\vspace*{10mm}

\begin{abstract}
\medskip
\noindent
We present predictions for $b$-quark production in photoprodcution 
and compare with experimental data from HERA. Our theoretical 
predictions are obtained at next-to-leading-order in the 
general-mass variable-flavor-number scheme, an approach which 
takes into account the finite mass of the $b$ quarks. We use 
realistic evolved nonperturbative fragmentation functions obtained 
from fits to $e^+e^-$ data. We find in general good agreement of 
data with both the GM-VFNS and the FFNS calculations, while the 
more precise ZEUS data seem to prefer the GM-VFNS predictions. 
\end{abstract}

\clearpage

\section{Introduction}

The investigation of heavy-quark (bottom or charm) production in $ep$ 
collisions is a useful test of perturbative Quantum Chromodynamics 
(QCD) since the heavy quark mass provides a hard scale that allows 
to perform calculations within perturbation theory. At leading order 
(LO), boson-gluon fusion, $\gamma g \rightarrow b \bar{b}$ is the 
dominant process for bottom-quark production. When the negative squared 
four-momentum of the initial and final electron, $Q^2$, is small, 
the process $ep \rightarrow e bX$ is treated as photoproduction, 
in which a quasi-real photon emitted by the initial electron interacts 
with the partons in the proton (direct contribution). For $Q^2 \simeq 
0$, in addition to the direct contributions, there are also 
resolved contributions, where the incoming photon changes into an 
initial quark or gluon, which interact with partons from the incoming 
proton. 

Bottom photoproduction has been measured using several different 
methods by the H1 \cite{Aaron:2012cj} and ZEUS \cite{ZEUS:2011aa} 
collaborations at HERA. In most of the measurements, the cross 
section was obtained using semi-leptonic decays into muons or 
electrons. All these measurements have been compared with predictions 
based on the fixed-flavor-number scheme (FFNS) \cite{Frixione:1995qc}. 
In this scheme, the bottom quark is generated in the hard 
scattering-process and appears only in the final state. Predictions 
for the photoproduction cross section are calculated at 
next-to-leading-order (NLO) taking the finite mass of the bottom 
quark explicitly into account. Reasonable agreement of the data 
with FFNS predictions has been obtained.

In the large-$p_T$ region, characterized by $p_T \gg m_b$, the 
so-called massless or zero-mass variable flavor-number-scheme 
(ZM-VFNS) \cite{Cacciari:1993mq,Kniehl:1995em,Cacciari:1995ej} 
is considered more appropriate. This is the conventional parton 
model approach, where the $b$ quark, considered massless like 
any other incoming parton, is also an incoming parton from the 
proton and, for the resolved contribution, from the incoming 
photon. This gives rise to additional contributions from 
hard-scattering subprocesses with $b$ quarks in the initial state, 
besides those with $u$, $d$, $s$ and $c$ quarks and the gluon ($g$). 
Predictions in this approach are reliable only in the large 
$p_T$-region where terms of the order $m_b^2/p_T^2$ are negligible. 
Calculations at NLO automatically resum leading and next-to-leading 
logarithmic terms $\propto \ln(p_T^2/m_b^2)$. At the same time, 
all non-logarithmic terms through $O(\alpha_s^2)$ are retained for 
$m_b = 0$. With the usual choice of renormalization ($\mu_R$) and 
factorization scales ($\mu_I$ and $\mu_F$ for initial and 
final-state factorization, resp.), $\mu_R = \mu_I = \mu_F = m_T = 
\sqrt{p_T^2+m_b^2}$, the results are dominated by contributions 
from the $b$-quark PDF of the proton down to $p_T = 0$.

The neglect of the terms of order $m_b^2/p_T^2$ in the hard scattering 
cross section is remedied in the general-mass variable-flavor-number 
scheme (GM-VFNS) \cite{Kniehl:2009ar}. This theoretical framework 
combines the FFN and the ZM-VFN schemes. The mass-dependent terms 
of the FFNS are added to the ZM-VFNS by applying subtractions in 
such a way that in the limit $p_T \to \infty$ the hard-scattering 
cross sections in the ZM-VFNS are recovered. However, with the 
conventional choice of scales indicated above, the results in 
the GM-VFNS are still dominated by the contributions of the 
$b$-quark PDF down to $p_T \simeq 0$. Therefore, there is no smooth 
transition from the GM-VFNS at large $p_T$ values to the FFNS in 
the small-$p_T$ range, and the GM-VFNS in its original definition 
fails to describe the small-$p_T$ cross sections.

The dominance of contributions with $b$-quarks in the initial state 
at small $p_T$-values is connected to the fact that this part is 
treated in the massless scheme. A calculation of the $b$-quark 
initiated subprocesses in a scheme with massive partons is not 
available for photo- and for hadroproduction\footnote[1]{
  For deep inelastic scattering heavy-quark initiated processes 
  at NLO with massive quarks have been considered in Ref.\ 
  \cite{Kretzer:1998ju}.
  }. In addition it would be necessary to correctly describe the 
effects of the finite bottom-quark mass $m_b$ in the DGLAP evolution. 
Attempts in this direction have been presented in 
\cite{deOliveira:2013tya}.

The cross section with massless partons is divergent for $p_T \to 
0$. For a realistic description it is therefore necessary to eliminate 
the contribution with massless $b$-quarks in the initial state in 
the small-$p_T$ region. For inclusive $B$-meson production in 
$pp~(p\bar{p})$ collisions this has been achieved in recent work 
together with Kniehl and Schienbein \cite{Kniehl:2015fla} by a 
suitable choice of scales $\mu_I$ and $\mu_F$. In this modified 
GM-VFNS the FFNS results with the exact $m_b$ dependence are 
recovered at small $p_T$. This way reasonable results were obtained 
for the inclusive $B$-meson production cross section measured with 
CDF at the Fermilab Tevatron and with LHCb at the CERN LHC.

It is the aim of this work to apply the same prescription for the 
choice of the factorization scales to $b$-quark photoproduction. 
For the resolved part the modified prescription works as described 
in \cite{Kniehl:2015fla} for hadron-hadron scattering. For the 
direct part a corresponding calculation has not been done yet. We 
shall do this here and compare our results to recent ZEUS and H1 
cross section data in order to study whether the modifications of 
the GM-VFNS can lead to a better agreement with presently available 
experimental data.

Cross sections for inclusive charmed-hadron photoproduction have 
been calculated in the past \cite{Kramer:2003jw,Kniehl:2009mh}, 
but without the scale modifications described in \cite{Kniehl:2015fla}. 
The corresponding experimental data are not available for very small 
$p_T$, where the modifications would be most clearly visible. For 
somewhat larger $p_T$, very good agreement of the various observables 
for inclusive $D^{*}$ photoproduction has been found 
\cite{Aaron:2011rdq}.

The outline of this paper is as follows. In Sect.\ 2 we introduce 
our strategy for the transition to the FFNS and compare our 
predictions with the ZEUS and H1 data. Our conclusions are presented 
in Sect.\ 3.

\section {Results and comparison with ZEUS and H1 data}

In this section, we shall consider a viable framework for inclusive 
$h_b$ production, where $h_b$ is any hadron containing a $b$-quark. 
Our aim is to unify prescriptions for the calculation of theoretical 
predictions at small and at large $p_T$-values. We shall compare 
with the differential cross section $d\sigma/dp_T$ measured by the 
ZEUS and the H1 collaborations at HERA. We take the $b$-quark pole 
mass to be $m_b=4.5$ GeV, evaluate $\alpha_s^{(n_f)}(\mu_R)$ at NLO 
with $n_f = 4$ and $\Lambda^{(4)}_{\overline{MS}} = 328$ MeV if 
$\mu_R < m_b$ and with $n_f=5$ and $\Lambda^{(5)}_{\overline{MS}} 
= 226$ MeV if $\mu_R > m_b$. For the proton PDFs we use the 
parametrization CTEQ6.6 \cite{Nadolsky:2008zw} while for the 
photon PDF we use AFG04\_BF \cite{Aurenche:2005da} as our default, but 
we will also consider other PDF choices for comparison.

\begin{figure*}
\begin{center}
\includegraphics[width=7.8cm]{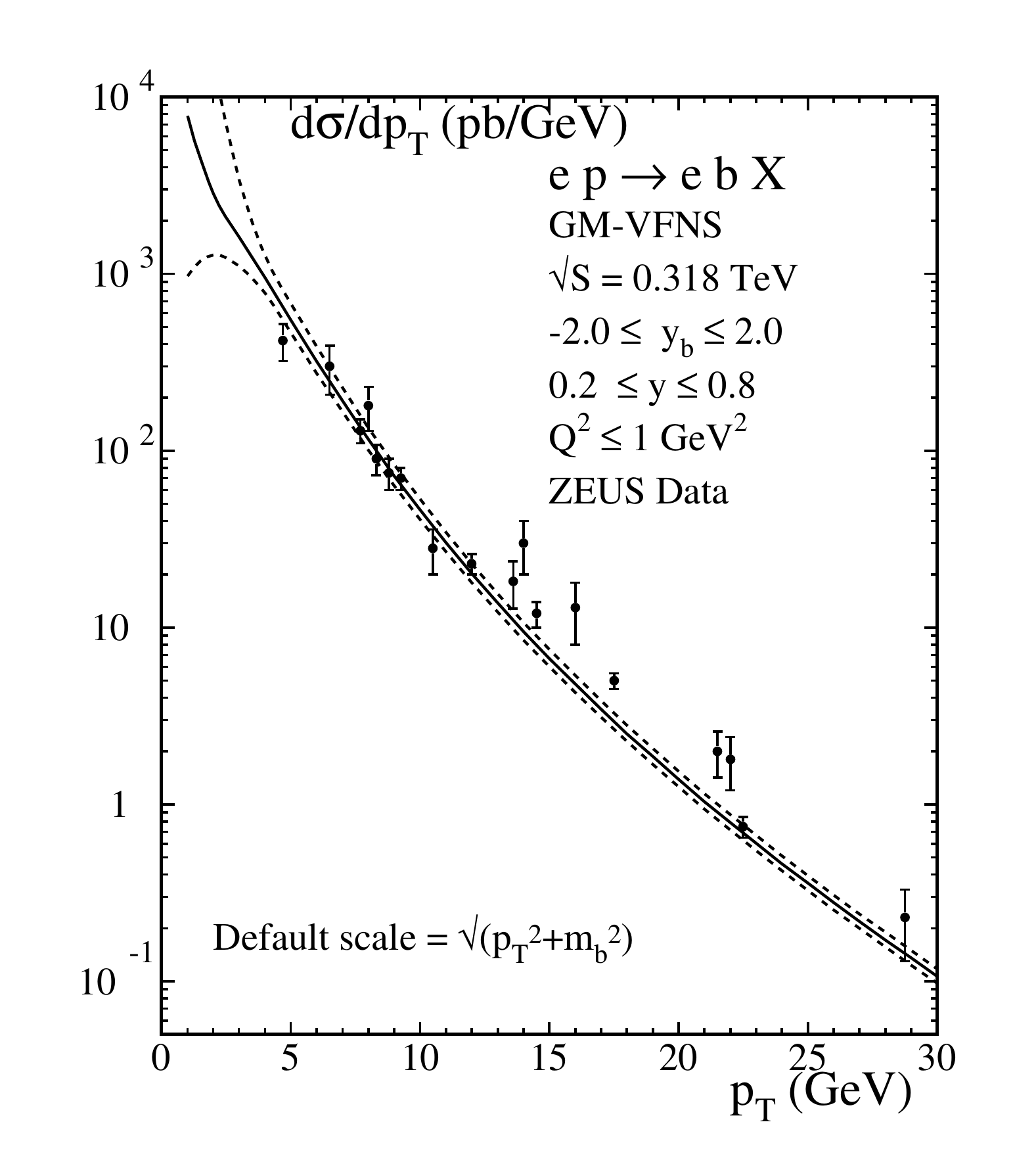}
\includegraphics[width=7.8cm]{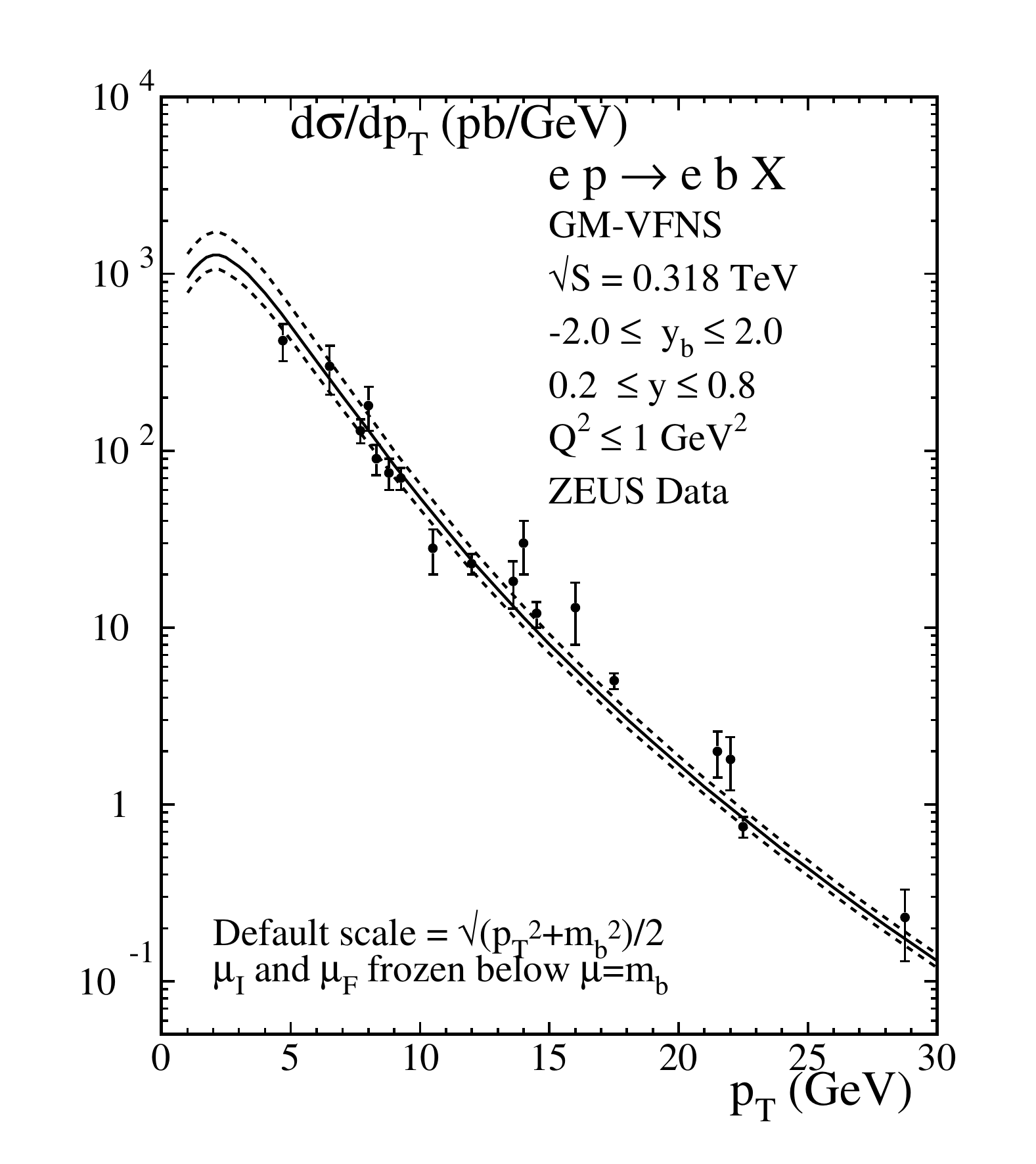}

\includegraphics[width=7.8cm]{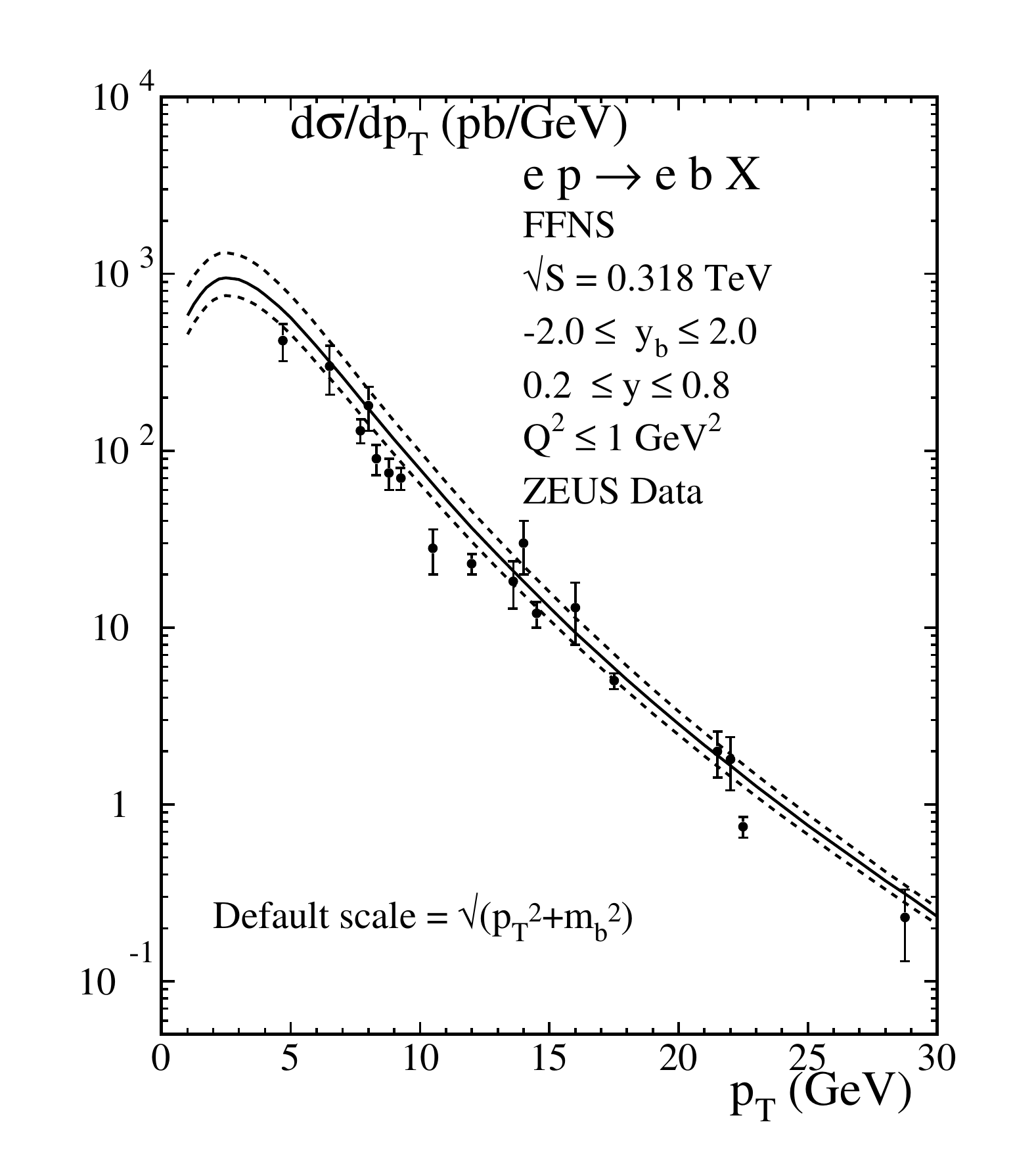}
\end{center}
\caption{
\label{fig:1}  
Differential cross section $d\sigma/dp_T$ for bottom-quark production 
as a function of the transverse momentum $p_T$ of the bottom quark 
compared with ZEUS data \cite{ZEUS:2011aa}. The upper left frame 
is for the original GM-VFNS choice of default scales $\mu_R = \mu_I 
= \mu_F = m_T$; the upper right frame shows results with the new 
default scale $\mu_R = m_T$, $\mu_I = \mu_F = m_T/2$. For comparison 
we also show in the lower panel results of the FFNS with 
$\mu_R = \mu_I = m_T$ (no FF). 
}
\end{figure*}

We start with results to be compared with the ZEUS data 
\cite{ZEUS:2011aa} shown in Fig.\ \ref{fig:1} together with our 
theoretical predictions. The data have been measured in various 
different analysis. Only five data points were reported in 
\cite{ZEUS:2011aa}. The other 13 points are from earlier publications. 
Their origin can be traced back from Fig.\ 8 of \cite{ZEUS:2011aa}. 
The ZEUS data are for the differential cross section for $b$-quark 
production as a function of $p_T$ (between 5 and 30 GeV) with the 
kinematic constraints $Q^2 < 1$ GeV$^2$ for the momentum transfer 
squared, $0.2 < y < 0.8$ for the inelasticity $y=Q^2/(x_{\rm Bj} S)$ 
(where $x_{\rm Bj}$ is the Bjorken scaling variable and $S$ the total 
energy in the center-of-mass reference frame) and $|y_b| < 2.0$ 
for the $b$-quark pseudo-rapidity. The data points at $p_T = 9.25$, 
$12.5$, $17.5$, $22.5$, $28.75$ GeV from \cite{ZEUS:2011aa} have in 
average smaller experimental errors than the earlier measurements 
since they are measured with the largest integrated luminosity of 
133~pb$^{-1}$. Note that not all points are consistent with each 
other, i.e.\ for some points the systematic errors are presumably 
larger than given in \cite{ZEUS:2011aa}. In most of the previous 
measurements of bottom photoproduction at HERA, the cross section 
was determined using semi-leptonic decays into muons or electrons. 
In \cite{ZEUS:2011aa} the $b$-quark photoproduction cross sections 
were determined for the production of jets containing $b$-quarks. 
The cross sections for the production of $b$-jets have been 
converted into cross sections for $b$-quark production with the 
help of the FMNR program \cite{Frixione:1993dg}. In Fig.\ 8 of 
Ref.\ \cite{ZEUS:2011aa} the data have been compared with predictions 
of the FFNS \cite{Frixione:1995qc}.

In Fig.\ \ref{fig:1} (upper left frame), we compare the ZEUS data 
with NLO predictions in the GM-VFNS. Our framework has been 
described in detail in Refs.\ \cite{Kramer:2003jw,Kniehl:2009mh} 
for photoproduction of $D^{*}$ mesons. The full line in Fig.\ 
\ref{fig:1} (upper left frame) shows the result for the old default 
choice of scales, $\mu_i = \xi_i m_T$ with $\xi_i = 1$ for $i = 
R$, $I$, $F$ where $R$, $I$, and $F$ denote renormalization, 
initial-state and final-state factorization scales. The dashed 
lines represent an estimate of the theoretical error by varying 
$\xi_i$ up and down by a factor of 2. 13 data points agree with 
the theoretical prediction within experimental and theoretical 
errors, 5 points lie outside this range. 

The central curve in Fig.\ \ref{fig:1} (upper left frame), as well 
as the curve for the upper error estimate, show the characteristic 
behavior of a monotonic increase of $d\sigma/dp_T$ in the limit 
$p_T \to 0$. This behavior is caused by the scale choice with 
$\xi_I \geq 1$. The curve for the lower error estimate with 
$\xi_I = 0.5$, however, shows a turn-over towards small $p_T$ with 
a maximum at $p_T \simeq 3$ GeV. This is the same characteristic 
behavior as in the FFNS \cite{Aaron:2012cj,ZEUS:2011aa} and is 
caused in our case by the fact that at around $p_T = 7.7$ GeV the 
factorization scales $\mu_i = \xi_i m_T = m_T/2$ fall below $m_b$ 
where the $b$-quark PDF and FF are zero. The same typical turn-over 
was also found in our previous study of inclusive $B$-meson 
production for $pp$ and $p\bar{p}$ collisions \cite{Kniehl:2015fla}.

Actually, the $b$-quark cross section is calculated from the 
$B$-meson cross section using an evolved fragmentation function 
(FF) for $b \to B$ taken from \cite{Kniehl:2008zza} and dividing 
by the branching fraction $Br(b\to B) = 0.40$. The effect of the 
evolved FF is visible at larger values of $p_T > 20$ GeV where 
the cross section $d\sigma/dp_T$ is smaller than without a FF. 
We emphasize that a consistent theoretical prescription of the 
GM-VFNS requires to include a FF describing the $b \to h_b$ 
transition since final state singularities have to be subtracted 
from the bare cross section in analogy to the calculation in the 
ZM-VFNS.

Unfortunately all the ZEUS measurements are for $p_T \geq 5$ GeV, 
so that any turn-over of the cross section $d\sigma/dp_T$ towards 
$p_T = 0$ is not measured. All data points agree reasonably 
well also with the GM-VFNS predictions with the original default 
scale $\mu_R = \mu_I = \mu_F = m_T$.

A monotonic increase of the cross section $d\sigma/dp_T$ for the 
production of a massive quark towards small $p_T$ is unphysical. 
From our previous work on $B$-meson production in $pp$ and $p\bar{p}$ 
collisions \cite{Kniehl:2015fla} we know how the choice of the 
factorization scales must be modified in order to obtain cross 
sections for $p_T \to 0$ which show the expected behavior. Therefore, 
as in \cite{Kniehl:2015fla}, we choose as the new default scales 
$\mu_R = m_T$ and $\mu_I = \mu_F = m_T/2$ such that both the 
$b$-quark PDF and the FF for the $b \to h_b$ transition vanish 
for $\mu_F < m_b$ already at non-zero $p_T$. With $\xi_R = 1$, 
$\xi_I = \xi_F = 0.5$ as the new default scale choice, we estimate 
the theoretical error in the usual way by varying the renormalization 
scale parameter $\mu_R$ by a factor 2 up and down about the default 
scale. We have not introduced an extra prescription to freeze 
$\mu_R$ below $m_b$ because the choice of $\mu_R$ is not related 
to switching off $b$-quark initiated subprocesses. The resulting 
error band is shown in Fig.\ \ref{fig:1} (upper right frame) together 
with the full curve for the new default choice and the ZEUS data 
\cite{ZEUS:2011aa}. The agreement with the data is only marginally 
better than in the upper left frame of Fig.\ \ref{fig:1}. Since the 
data have $p_T$ values larger than 5 GeV the turn-over of 
$d\sigma/dp_T$ towards $p_T \to 0$ is not tested by the ZEUS 
measurements. 

In the lower part of Fig.\ \ref{fig:1} we compare the ZEUS data 
with results obtained in the FFNS. Here we have fixed the scales 
to $\mu_R = \mu_I = m_T$. The FFNS prediction is only slightly 
below the GM-VFNS calculation at low $p_T$, but lies above it by 
about a factor of two at the largest $p_T$. The difference at large 
$p_T$ is mainly due to the fact that the FFNS calculation does not 
include an evolved FF. We observe that the FFNS calculation 
over-shoot the high-precision data points of ZEUS. 

\begin{figure*}
\begin{center}
\includegraphics[width=7.8cm]{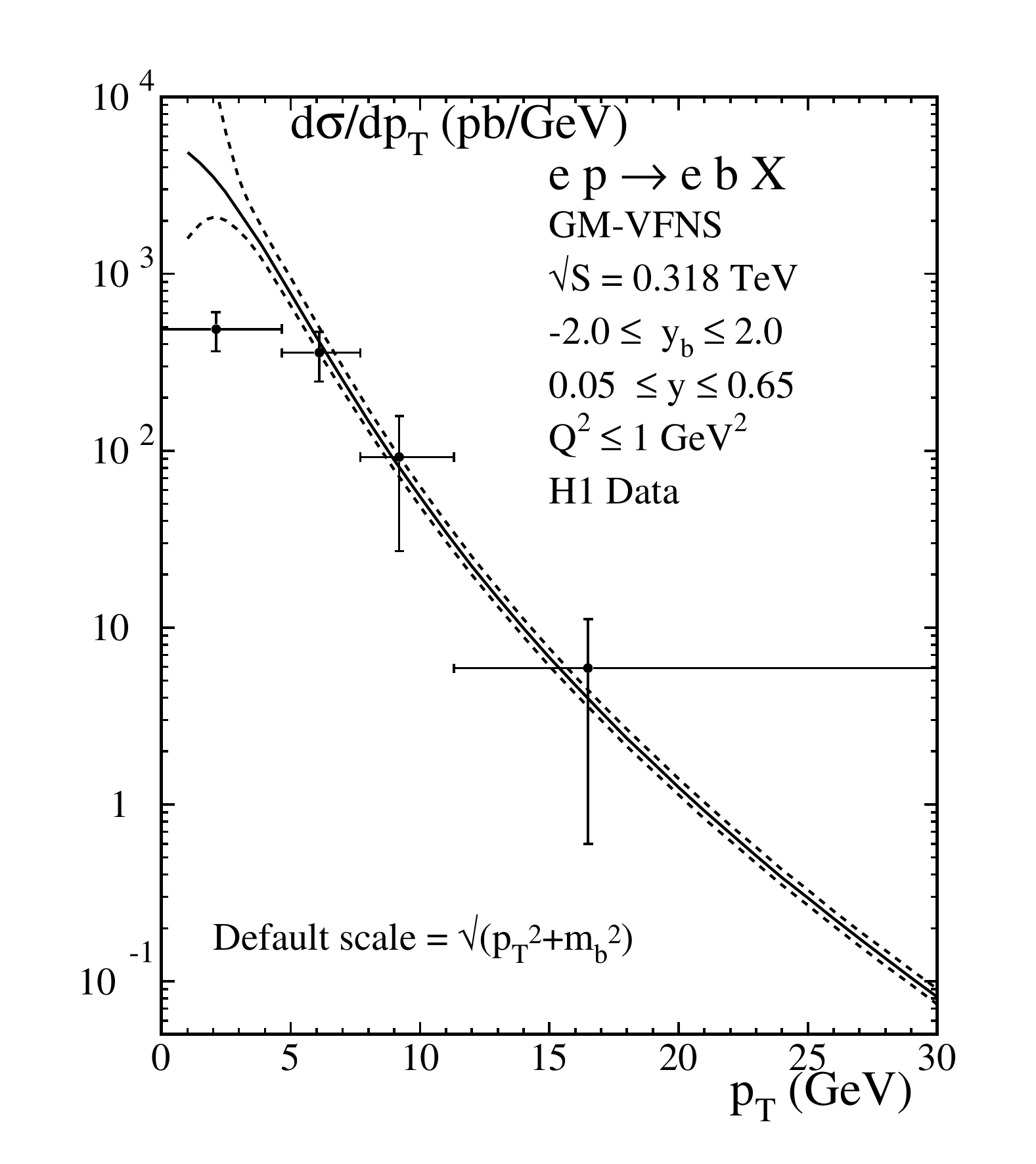}
\includegraphics[width=7.8cm]{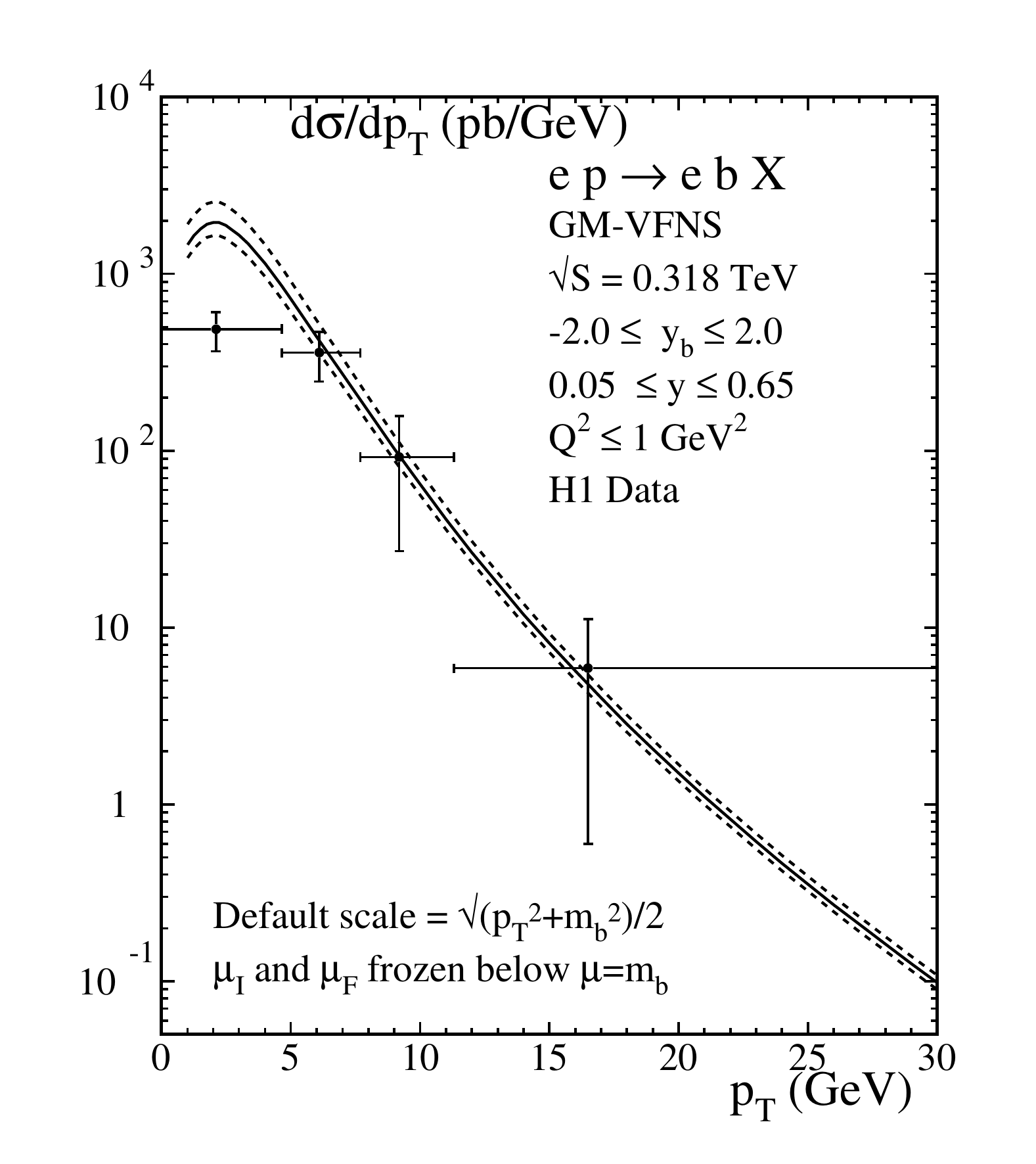}

\includegraphics[width=7.8cm]{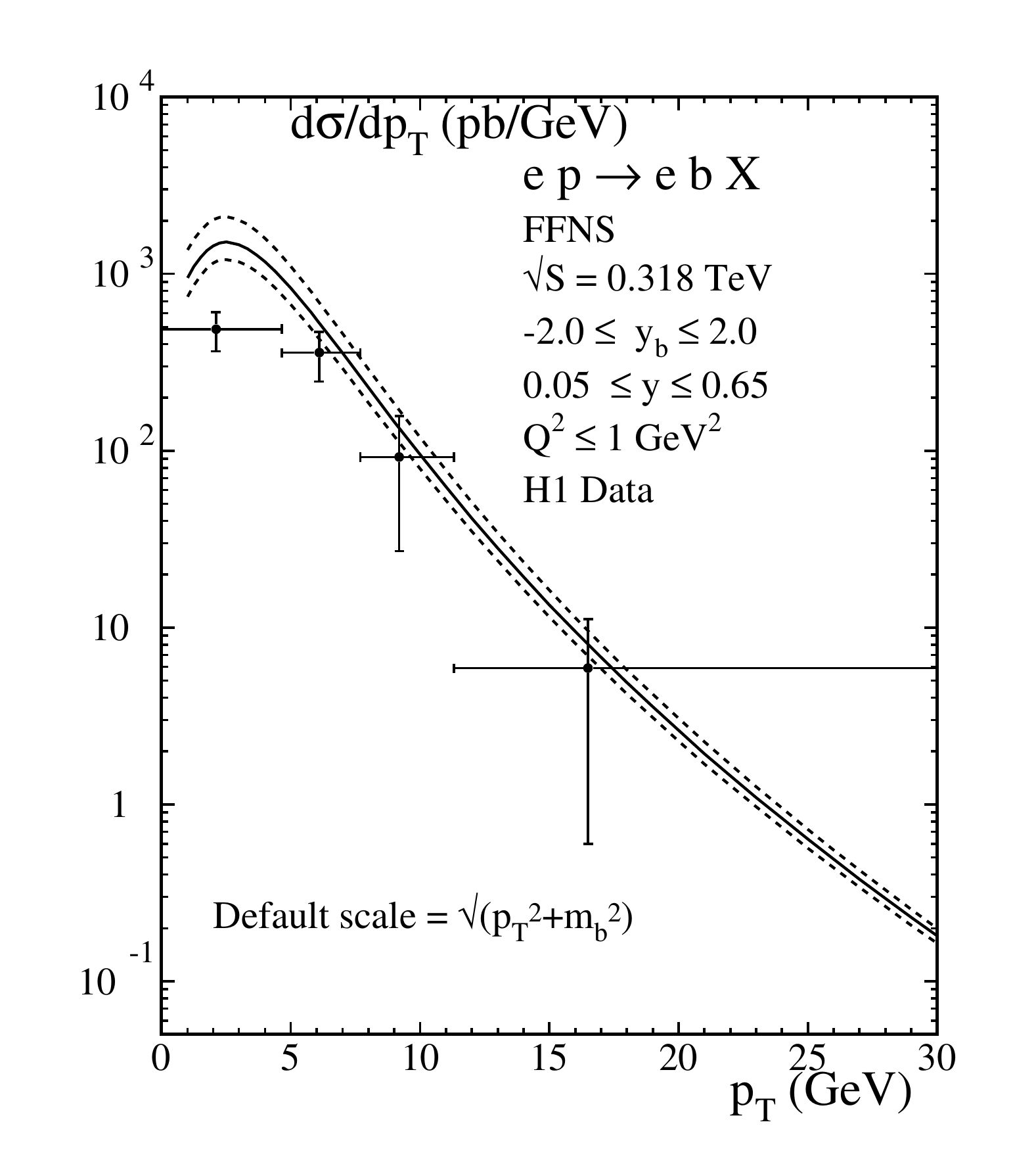}
\end{center}
\caption{
\label{fig:3} 
Differential cross section $d\sigma/dp_T$ for bottom photoproduction 
as a function of the transverse momentum $p_T$ compared to H1 
data \cite{Aaron:2012cj}. The upper left frame is for the original 
GM-VFNS choice of scale parameters $\mu_R = \mu_I = \mu_F = m_T$ 
and the upper right frame for the new default scales $\mu_R = m_T$, 
$\mu_I = \mu_F = m_T/2$. FFNS results with $\mu_R = \mu_I = m_T$ 
(no FF) are shown in the lower frame. 
}
\end{figure*}

Similar data are available from the H1 collaboration at HERA. The 
most recent publication \cite{Aaron:2012cj} describes results for 
photoproduction of $b$-quarks measured in the process $e p \to e b 
\bar{b} X$. The decay channel $b \bar{b} \to eeX$ is selected by 
identifying the semi-electronic decays of the $b$-quarks. The 
production cross section is measured in the kinematic range where 
the photon virtuality is small, $Q^2 \leq 1$ GeV$^2$, the inelasticity 
is restricted to the range $0.05 \leq y \leq 0.65$ and the pseudorapidity 
of the $b$-quarks is in the range $|y_b| \leq 2$. The measured cross 
sections are converted into single-inclusive $b$-quark production 
cross sections $d\sigma/dp_T$ for four $p_T$ bins in the range $0 
\leq p_T \leq 30$ GeV. We have calculated the $ep \to e b X$ cross 
section $d\sigma/dp_T$ as a function of $p_T$ in this $p_T$ range 
and with the same kinematic constraints as in \cite{Aaron:2012cj} 
in the original GM-VFNS with default scales $\xi_R = \xi_I = \xi_F 
= 1$ and with the new default scales $\xi_R = 1$, $\xi_I = \xi_F = 
0.5$. The results, together with scale variations as above to obtain 
an error estimate, are shown in Fig.\ \ref{fig:3} together with the 
H1 data. The upper left panel of Fig.\ \ref{fig:3} shows results for 
the original scale setting, the upper right frame is for the new 
scale choice. The experimental data for $p_T > 5$ GeV agree with 
the predictions in both schemes. Only the cross section $d\sigma/dp_T$ 
in the lowest-$p_T$ bin, $0 \leq p_T \leq 4.65$ GeV, is lower than 
the predictions in both cases. In this $p_T$-bin the measured cross 
section is smaller than the theory prediction by approximately a 
factor of 6 in the upper left frame of Fig.\ \ref{fig:3} and by a 
factor 3 to 4 in the upper right frame of Fig.\ \ref{fig:3}. Although 
the $p_T$-dependence of the differential cross section $d\sigma/dp_T$ 
exhibits the expected turn-over at $p_T \simeq 2$ GeV if the new 
default scales are chosen, the predicted cross section is still 
larger than the experimental data, even when taking into account 
the experimental uncertainty. 

A comparison of the data with the FFNS prediction is shown in 
the lower frame of Fig.\ \ref{fig:3}, again choosing scales as 
$\mu_R = \mu_I = m_T$ and without folding with a FF. As above, 
the FFNS results are slightly below those of the GM-VFNS at low 
$p_T$, but above at large $p_T$. We note that the NLO corrections 
in the FFNS are positive and lead to an increase of the cross 
section by about a factor of 1.5 at low $p_T$. The H1 data with 
its relatively large errors are unable to discriminate between 
the two schemes. We should remark that the H1 data are presented 
as a function of the average transverse momentum of two produced 
bottom quarks, while our calculation is for the one-particle 
inclusive cross section where the independent kinematic variable 
is the transverse momentum of one observed $b$-quark. This may 
be the reason that the theoretical predictions shown in 
\cite{Aaron:2012cj} do not agree with our FFNS evaluation.   

\begin{figure}[t]
\begin{center}
\includegraphics[width=8cm]{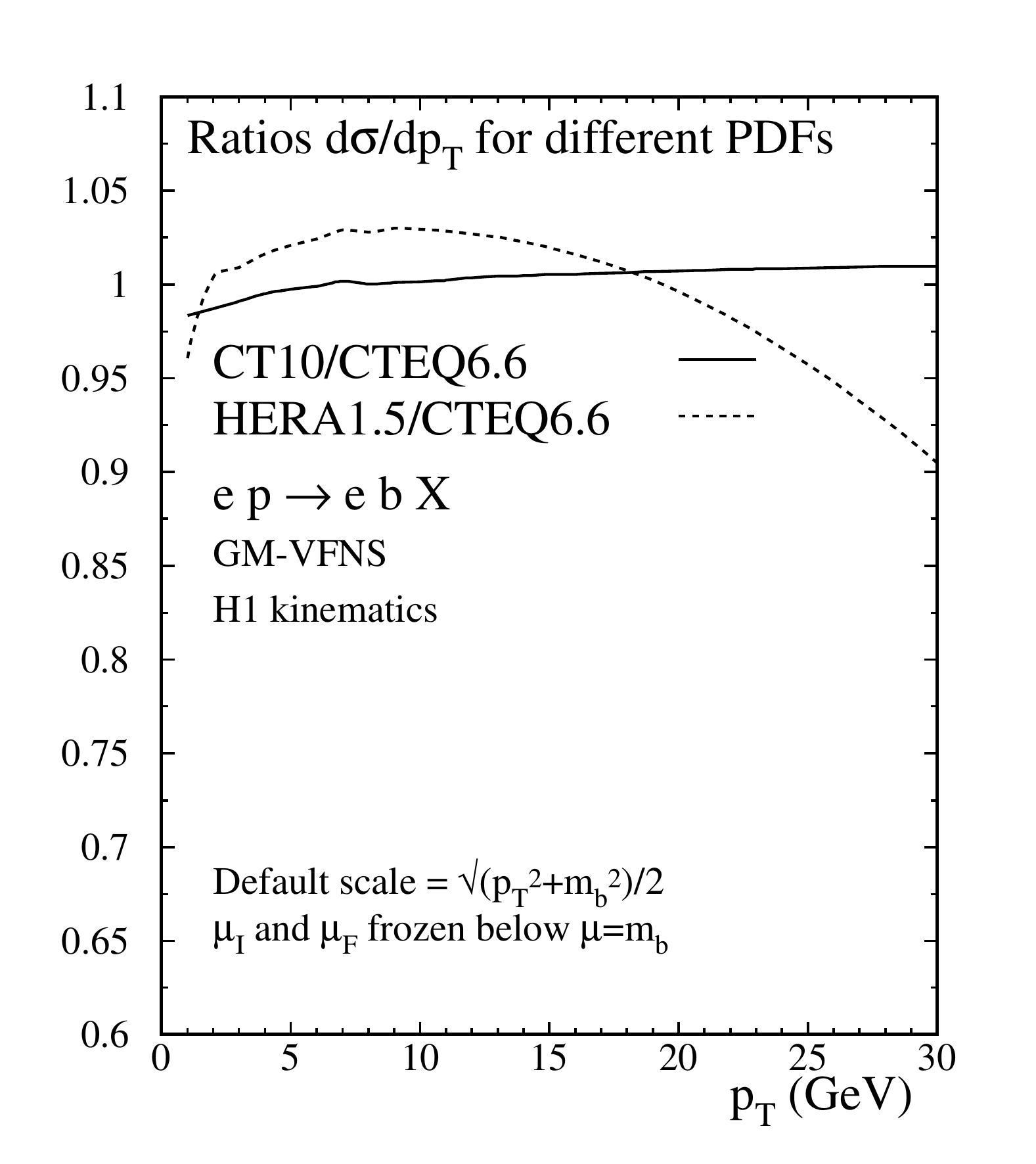} 
\includegraphics[width=8cm]{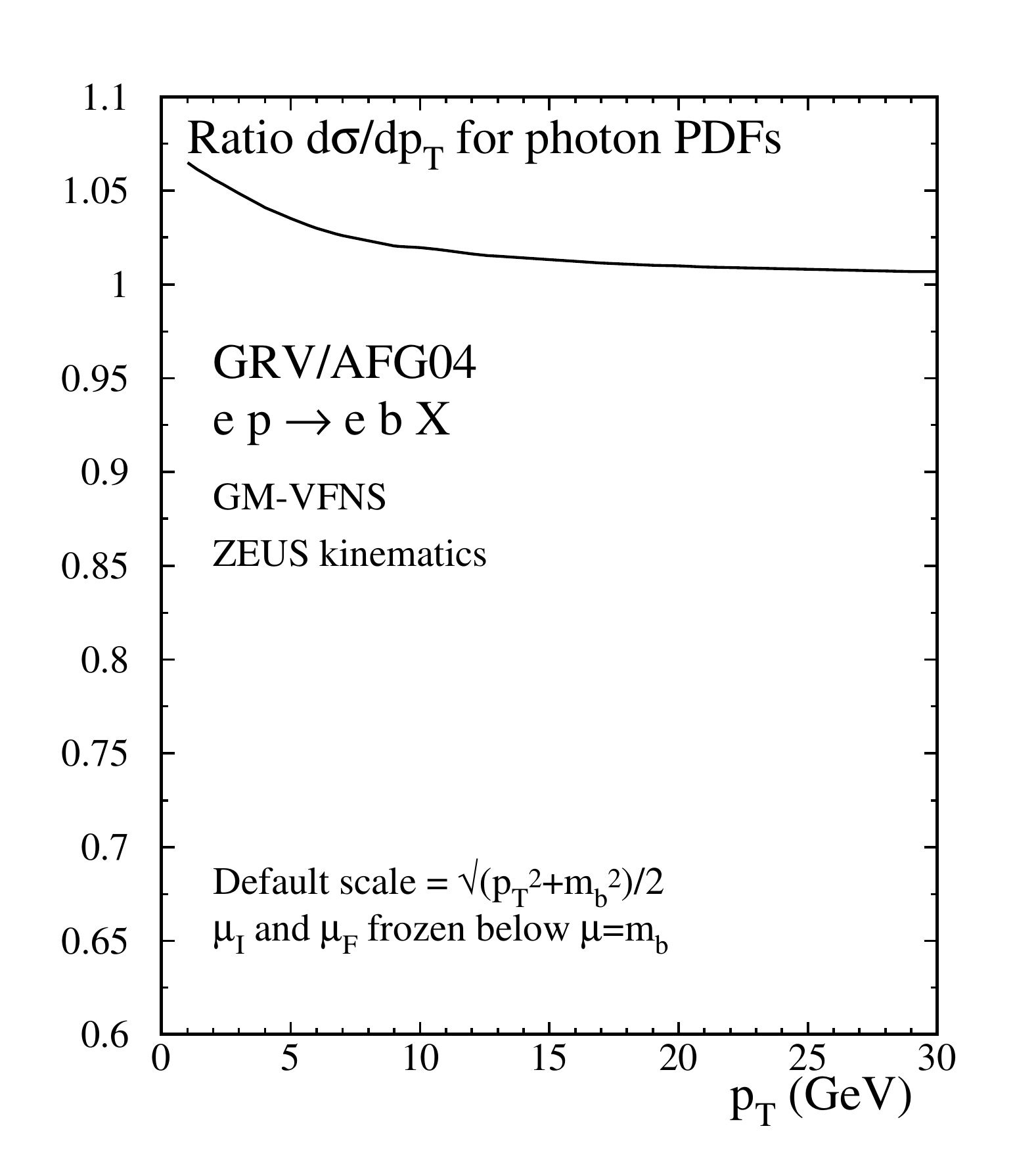}
\end{center}
\caption{
\label{fig:2} 
Left frame: ratios of the differential cross sections $d\sigma/dp_T$ 
for bottom photoproduction using different proton PDF parametrizations 
with respect to the default CTEQ6.6 \cite{Nadolsky:2008zw} PDF. The 
full curve is for CT10 \cite{Lai:2010vv}, the dashed curve for 
HERAPDF1.5 \cite{Alekhin:2014irh}. 
\newline
Right frame: ratio of the differential cross section $d\sigma/dp_T$ 
for bottom photoproduction as a function the transverse momentum 
$p_T$ for the photon PDF GRV HO \cite{Gluck:1991jc,Gluck:1991ee} 
with respect to the photon PDF AFG04 \cite{Aurenche:2005da}. 
In both figures, the kinematic constraints of the ZEUS data and 
the new default scales $\mu_R = m_T$, $\mu_I = \mu_F = m_T/2$ 
have been used.
}
\end{figure}

In order to study the influence of the choice of the proton PDF 
we present also results for two other PDFs: CT10 \cite{Lai:2010vv} 
and HERAPDF1.5 (NLO) \cite{Alekhin:2014irh}. The results are shown in 
Fig.\ \ref{fig:2} (left frame) as ratios to the prediction obtained 
for the default choice CTEQ6.6. The differences to CTEQ6.6 are very 
small, in particular for CT10, as to be expected. For HERAPDF1.5 
the difference is somewhat larger. In the range $2 < p_T < 20$ GeV 
the cross section with HERAPDF1.5 is larger than for CTEQ6.6 
and for larger $p_T$ values, $p_T > 20$ GeV, the ratio is smaller 
than 1 and decreases towards 0.9 for $p_T = 30$ GeV.

We include also a comparison with two different choices of the 
photon PDFs, shown in the right frame of Fig.\ \ref{fig:2}. Again, 
we present the results as a ratio of cross sections and choose 
the GRV HO \cite{Gluck:1991jc,Gluck:1991ee} photon PDF as an 
alternative to the AFG04 parametrization. The cross section ratio 
is very close to 1. Only for the smallest considered $p_T$ value 
the ratio reaches values slightly above $1.06$.

\begin{figure*}[t]
\begin{center}
\includegraphics[width=8cm]{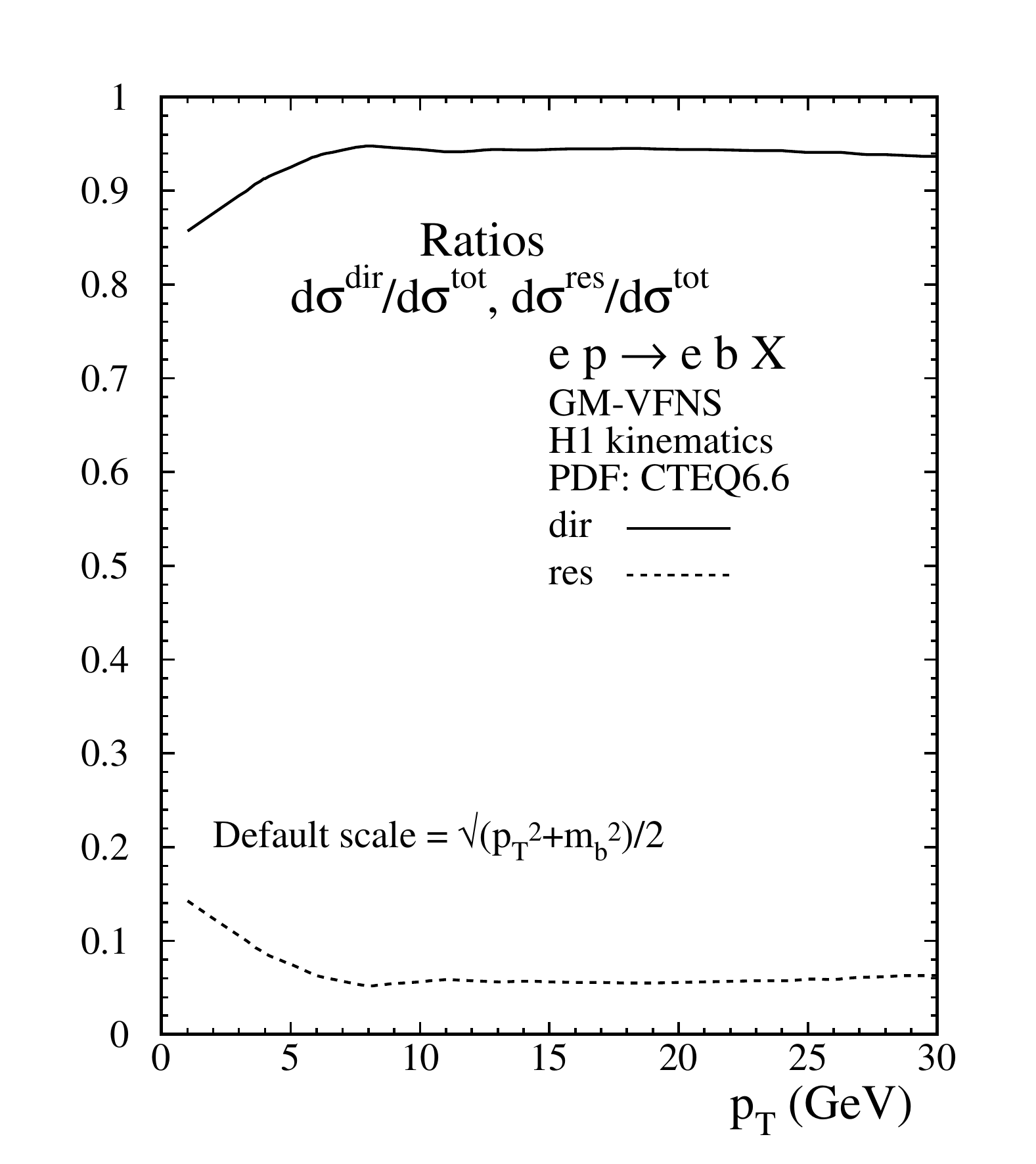}
\end{center}
\caption{
\label{fig:4} 
Ratios of $d\sigma^{\rm dir}/dp_T$ and $d\sigma^{\rm res}/dp_T$ to 
the total cross section $d\sigma^{\rm tot}/dp_T$ as a function of 
the transverse momentum $p_T$ of the $b$-quark. 
}
\end{figure*}

As is well-known, a complete calculation of the photoproduction 
cross section has to take into account two parts, the direct 
contribution and the resolved contribution. It is interesting 
to study whether both parts show the same dependence on $p_T$. 
To see this in some detail we have calculated these two parts 
separately, denoted $d\sigma^{\rm dir}/dp_T$ and $d\sigma^{\rm res} 
/ dp_T$. Results for the H1 kinematic conditions are shown in 
Fig.\ \ref{fig:4}. For the sake of a more clear visibility we 
display the ratios of $d\sigma^{x}/dp_T$ and $d\sigma^{\rm tot}/dp_T$ 
where $x = {\rm dir}$, res and $d\sigma^{\rm tot}/dp_T = 
d\sigma^{\rm dir}/dp_T + d\sigma^{\rm res}/dp_T$ is the sum of the 
direct and resolved contributions as shown Fig.\ \ref{fig:3} 
(upper right frame). From Fig.\ \ref{fig:4} we see that the 
resolved cross section varies between $6\%$ and $15\%$ for 
$p_T < 5$ GeV, while it stays near $6\%$ for the larger $p_T$ 
values. This means that even when the resolved part would vanish 
for all $p_T$ the direct part would change by less than $15\%$. 
This is too small to be identified unambiguously in the data. In 
particular, a different $p_T$ dependence of the direct part can 
not be made responsible for the low cross section measured by H1 
at $p_T=2$ GeV.

%


\section {Conclusions}

We have presented predictions for $b$-quark photoproduction 
at HERA at next-to-leading order in the general-mass 
variable-flavor-number scheme. In contrast to previous 
calculations, we have fixed the factorization scale parameters 
in such a way that $b$-quark initiated contributions are 
eliminated at low transverse momenta. With this new prescription 
we revocer the typical low-$p_T$ behaviour of a fixed-flavor-number 
scheme. The comparison with experimental data from the ZEUS 
collaboration shows reasonable agreement over the whole range of 
transverse momenta, while we observe a discrepancy in the 
lowest-$p_T$ bin of the H1 measurement.


\end{document}